\documentstyle[aps,prl,epsf]{revtex}

\newcommand{\be}{\begin{equation}}
\newcommand{\ee}{\end{equation}}
\newcommand{\bea}{\begin{eqnarray}}
\newcommand{\eea}{\end{eqnarray}}
\newcommand{\bdm}{\begin{displaymath}}
\newcommand{\edm}{\end{displaymath}}

\newcommand{\fourF}{F^{(4)}}
\newcommand{\fourA}{A^{(4)}}
\newcommand{\fourR}{R^{(4)}}
\newcommand{\fourD}{D^{(4)}}
\newcommand{\fourast}{\ast^{(4)}}
\newcommand{\we}{\wedge}

\newcommand{\Trace}[1]{\mbox{Tr} \left\{ #1 \right\}}
\newcommand{\idid}{1 \! \! 1}

\newcommand{\RR}{\mbox{$I \! \! R$}}
\newcommand{\gtens}{\mbox{\boldmath $g$}}
\newcommand{\fourg}{{\mbox{\boldmath $g$}}^{(4)}}
\newcommand{\bx}{\mbox{\boldmath $x$}}
\newcommand{\by}{\mbox{\boldmath $y$}}
\newcommand{\bz}{\mbox{\boldmath $z$}}
\newcommand{\bv}{\mbox{\boldmath $v$}}

\newcommand{\bbJJ}{\mbox{\boldmath $J$}}
\newcommand{\bbAA}{\mbox{\boldmath $A$}}
\newcommand{\bbBB}{\mbox{\boldmath $B$}}
\newcommand{\bbPP}{\mbox{\boldmath $P$}}

\begin{document}

\title{Stationary perturbations and infinitesimal rotations of 
static Einstein-Yang-Mills configurations with bosonic matter}

\author{Othmar Brodbeck and Markus Heusler}

\address{Institute for Theoretical Physics \\
The University of Zurich \\
CH--8057 Zurich, Switzerland}

\date{\today}

\maketitle

\begin{abstract}
Using the Kaluza-Klein structure of stationary spacetimes,
a framework for analyzing stationary perturbations of static
Einstein-Yang-Mills configurations with bosonic matter fields
is presented. It is shown that the perturbations giving
rise to non-vanishing ADM angular momentum are governed by a
self-adjoint system of equations for a set of gauge invariant 
scalar amplitudes. The method is illustrated for SU(2) gauge 
fields, coupled to a Higgs doublet or a Higgs triplet. It is 
argued that slowly rotating black holes arise generically in 
self-gravitating non-Abelian gauge theories with bosonic matter, 
whereas, in general, soliton solutions do not have rotating 
counterparts. 
\end{abstract}


\section{Introduction}

In the presence of a (stationary) Killing symmetry, the 
Einstein-Maxwell (EM) equations reduce to a $\sigma$-model 
coupled to three-dimensional gravity \cite{Neugebauer69}. 
This property is, in fact, shared by a large class of theories
with scalar fields and {\it Abelian\/} vector fields (see 
\cite{Breitenlohneretal88} for a classification and 
\cite{GaltsovLetelier97}, \cite{MH97} for some recent 
applications and references). If spacetime admits an 
additional (axial) Killing symmetry, then the $\sigma$-model 
structure gives rise to total integrability of the field 
equations, provided that the target space is a symmetric 
space. This has been known for quite some time for the EM
system \cite{Geroch72} and was recently demonstrated by 
Gal'tsov for EM-dilaton-axion models \cite{Galtsov95}.

Since {\it scalar} magnetic potentials fail to exist in 
{\it non-Abelian\/} gauge theories, the $\sigma$-model 
structure -- and, in particular, the property of 
integrability -- are spoiled for self-gravitating Yang-Mills 
fields. Moreover, the circularity theorem \cite{Carter70} 
(which guarantees that spacetime admits a foliation by 
two-surfaces orthogonal to the integral trajectories of the 
two Killing fields) does not extend to the Einstein-Yang-Mills 
(EYM) system \cite{MH96JR} (see also \cite{MH96BC}). The 
familiar Papapetrou metric \cite{Papapetrou53} does, therefore, 
not take account of {\it all\/} stationary and axisymmetric 
degrees of freedom of the EYM equations.

In view of these problems, an analytic approach to the full 
EYM equations with two Killing fields is likely to be 
extremely difficult. Motivated by recent work of Straumann 
and Volkov \cite{StraumannVolkov97}, we pursue a more modest 
aim in this paper, that is, we 
consider stationary {\it deviations} 
of static EYM configurations with bosonic matter fields. 
For the pure SU(2) EYM system, Straumann and Volkov 
\cite{StraumannVolkov97} were able to reduce the relevant 
perturbation equations to a three-dimensional set. In this 
paper, we present a systematic investigation of stationary 
perturbations, which reveals that the decoupling of a specific 
set of perturbation amplitudes is a {\it general feature\/} 
of a large class of bosonic matter fields coupled to the EYM 
system with an arbitrary gauge group. We argue that 
stationary perturbations are most appropriately
handled by means of a ``$3+1$'' -- rather than 
a ``$2+2$'' -- decomposition of spacetime. (This does, 
in particular, avoid the circularity issue, since the 
metric is not required to be axially symmetric in the first 
place.) Hence, we use the Kaluza-Klein (KK) structure of a 
stationary spacetime to analyze {\it arbitrary stationary\/} 
perturbations of static configurations. Within this approach, 
the {\it non-static} deviations are encoded in the KK 
connection, which is related to (the dual of) the 
twist of the stationary Killing field.

The KK reduction of the Einstein-Hilbert action yields a 
three-dimensional gravitational theory coupled to the KK 
scalar field and the the KK connection \cite{Ehlers59}. The 
latter is described by a gauge potential, which enters the 
effective action only quadratically and only via the field 
strength. Using a suitable KK decomposition of the YM gauge 
potential, these features are preserved if gauge fields and 
additional bosonic matter are coupled to gravity. More 
precisely, it turns out that the linear terms in the KK 
connection enter the reduced action via a (non-minimal)
coupling to the electric components of the YM field. These 
observations imply the following two conclusions: (i) The 
stationary EYM equations (coupled to bosonic fields) admit a 
generalized scalar twist potential, and (ii) the non-static, 
non-magnetic deviations of a static, purely magnetic solution 
to the EYM equations form a consistent subset of all stationary 
perturbations. Moreover, it is exactly this subset of 
perturbations, henceforth called {\it purely stationary\/} 
perturbations, which gives rise to a non-vanishing ADM 
angular momentum.

By virtue of the crucial features (i) and (ii), the relevant 
perturbations (as far as angular momentum is concerned) of a 
static, purely magnetic EYM-Higgs configuration form a formally 
{\it self-adjoint\/} system for a set of {\it gauge invariant\/} 
scalar amplitudes. For a spherically symmetric SU(2) background, 
these amplitudes, consisting of the generalized 
twist potential and the (Lie algebra valued) electric YM 
potential, can be expanded in terms of ``isospin'' harmonics,
${C}^{\,\ell}_{j\/m}$. Since only $j=1$ contributes to the 
ADM angular momentum, one finally obtains a standard 
Sturm-Liouville problem for three radial functions. For the 
twist channel one has $j=\ell=1$, whereas the orbital angular 
momenta in the two YM channels are $\ell=0$ and $\ell=2$. The 
Higgs fields enter the perturbation equations only via a 
background potential, which gives mass to either one (triplet) 
or both (doublet) YM perturbations.

For a stationary background, the horizon is a regular singular 
point of the perturbation equations, which admit four acceptable 
solutions, whereas the corresponding number is three in the 
asymptotic regime. The fact that the perturbation equations admit a 
six-dimensional fundamental system then yields the conclusion 
that slowly rotating black hole solutions to the EYM-Higgs 
equations do exist. The corresponding solutions for the
pure EYM system were recently discovered by Volkov and 
Straumann \cite{StraumannVolkov97}, who also argued that
these configurations cannot be electrically 
neutral. The perturbation equations show that the coupling of 
isospin and orbital momentum, which is responsible for 
the ``charging up'' due to rotation, does not need to be 
effective if bosonic matter is coupled to the EYM equations.

For solitonic background solutions the origin is a regular 
singular point of the perturbation equations. The number of 
physically acceptable modes at the center is, however, not 
sufficiently large to allow for ``generic'' rotational degrees 
of freedom of self-gravitating bosonic matter coupled to
non-Abelian gauge fields.

\section{Kaluza-Klein reduction}

We consider the action for self-gravitating non-Abelian gauge 
fields coupled to bosonic matter,
\be 
{\cal S} \, = \, -\frac{1}{16 \pi G} \, \int \left[ {\cal L}_{G} \, + 
\, \kappa \left( {\cal L}_{Y\!M} \, + \, {\cal L}_{B} \right) 
\right] \, ,
\label{action}
\ee
where $\kappa = 8 \pi G/g^2$, $G$ is Newton's constant, 
and $g$ is the gauge coupling. The four-forms
${\cal L}_{G}$ and ${\cal L}_{Y\!M}$ are the 
Einstein-Hilbert and the YM Lagrangians, respectively,
\be
{\cal L}_{G} \, = \, \fourast \fourR \, , \qquad 
{\cal L}_{Y\!M} \, = \, 2 \, 
\Trace{\fourF \we \fourast \fourF} \, .
\label{EYM-Lagr}
\ee
Here, $\fourR$ and $\fourast$ denote the Ricci scalar and 
the Hodge dual with respect to the spacetime metric $\fourg$.
The one-form $\fourA$ is the Lie algebra valued YM gauge 
potential with field strength 
$\fourF = d \fourA + \fourA \we \fourA$. For the bosonic 
matter we shall, for instance, consider a Higgs field $H$ 
[with potential $P(H)$] which transforms according to some 
representation $U$ of the gauge group, 
$\fourD H = d H + U_{\star}(\fourA) H$. In particular,
\be
{\cal L}_{B} \, = \, - 2 \,
\Trace{\left(\fourD H \right)^{\dagger} \we \fourast \fourD H} 
\;  -  \fourast P(H) \, ,
\label{Higgs-Lagr}
\ee
for a Higgs doublet or a triplet in matrix representation
[see also Eq. (\ref{back-H})]. 

Our first aim is to perform the KK reduction of the above 
action (\ref{action}). At least locally, a stationary spacetime 
$(M,\fourg)$ [with Killing field $\partial_t$ and corresponding
one-form $k = -\sigma(dt+a)$] has the structure 
$\RR \times \Sigma$ and admits a metric of KK type, 
\be
\fourg \, = \, -\sigma \, (dt + a) \otimes (dt + a) 
\, + \, \sigma^{-1} \gtens \, .
\label{KK-metric}
\ee
Here, $\sigma$ and $a$ are, respectively, a scalar field and a 
one-form on the three-dimensional Riemannian space
$(\Sigma, \gtens)$. Under coordinate transformations the 
one-form $a$ transforms like an Abelian gauge potential. The 
corresponding field strength, $da$, is proportional to the dual 
of the twist one-form, 
$\omega \equiv \frac{1}{2} \fourast(k \we dk)$ $=$ 
$-\frac{1}{2} \sigma^2 \ast da$. (Here and in the following
$\ast$ denotes the Hodge dual with respect to the Riemannian 
metric $\gtens$.) The canonical decomposition of the gauge 
field $\fourA$ in terms of a stationary function $\phi$ and a 
stationary one-form $A$ (both Lie algebra-valued) on 
$(\Sigma, \gtens)$ is
\be
\fourA \, = \, \phi \, (dt + a) \, + \, A \, .
\label{KK-gaugefield}
\ee
In the following it will be crucial that $\fourA$ is decomposed 
with respect to the orthonormal tetrad field
$\theta^{0} = \sqrt{\sigma}(dt + a)$ (rather than 
$\sqrt{\sigma}dt$). The reduction of the Einstein-Hilbert action 
with respect to the stationary metric (\ref{KK-metric}) gives 
$\int {\cal L}_{G} = \int (dt \we L_{G})$, where the
three-form $L_{G}$ is the Lagrangian for the KK scalar field 
$\sigma$ and the Abelian gauge field $a$, effectively coupled 
to $3$-dimensional gravity. Up to an exact differential, one finds
\be
L_{G} \, = \,
\ast R^{(g)} - \frac{1}{2 \sigma^{2}} \, 
d\sigma \we \ast d\sigma + \frac{\sigma^{2}}{2} \,
da \we \ast da \, .
\label{L-G}
\ee
The dimensional reduction of the YM action yields 
an effective YM-Higgs theory, with effective Higgs field 
$\phi$ and YM field strength $F \equiv dA + A \we A$. With
$\int {\cal L}_{Y\!M} = \int (dt \we L_{Y\!M})$ one has
\be
L_{Y\!M} \, = \, 2 \, \Trace{ \sigma \,
(F + \phi \, da) \we \ast (F + \phi \, da) \, - \, 
\frac{1}{\sigma} \, D \phi \we \ast D \phi } \, ,
\label{L-YM}
\ee
where $D$ denotes the (gauge) covariant exterior derivative 
with respect to the one-form $A$ on $\Sigma$. Introducing a 
field strength vector with components $da$ and $F$, the 
above formulas imply that the stationary EYM system reduces 
to a three-dimensional EYM theory which is non-minimally coupled 
to a two-component vector of scalar fields (comprising combinations 
of the KK scalar $\sigma$ and the YM scalar $\phi$).
Finally, the evaluation of the Higgs action with respect to the 
gauge potential (\ref{KK-gaugefield}) results in an additional 
potential term, involving the coupling between the actual Higgs field 
$H$ and the effective Higgs field $\phi$:
\be
L_{B} \, = \, - 2 \, \Trace{
\left( D H \right)^{\dagger} \we \ast D H \, - 
\, \frac{1}{\sigma^{2}} 
\, \left( U_{\star}(\phi) H \right)^{\dagger} 
\we \ast U_{\star}(\phi) H} 
\, - \, \frac{1}{\sigma} \ast P[H] \, .
\label{L-H}
\ee

The vacuum Einstein equations are obtained from variations
of $\int \! L_{G}$ with respect to $\gtens$, $\sigma$ and $a$.
Since $L_{G}$ is a quadratic expression in terms of $da$,
both the effective three-dimensional Einstein equation for 
$\gtens$ and the equation for $\sigma$ contain no linear 
terms in $da$. In the presence of YM and Higgs fields this 
property generalizes in the sense that the effective action 
continues to be quadratic in combinations of $da$ and $\phi$. 
Hence, the only equations which contain linear terms in $da$ 
and/or $\phi$ are those which are obtained from variations 
of the effective action, $\int_{\Sigma} 
\left[ L_G + \kappa \left( L_{Y\!M}+L_B \right) \right]$, 
with respect to these quantities:
\be
d \ast \left[ \sigma^{2} \, da \, + \, 
4 \/ \kappa \, \sigma \, \Trace{ \phi \, ( F + \phi \/ da ) }
\right] \, = \, 0 \, ,
\label{eq-a}
\ee
\be
D \ast \left[ \sigma^{-1} D \,\phi \right] 
\, + \, \sigma \, da \we \ast (F + \phi \/ da)  \, = \,
\sigma^{-2} \ast J_B(\phi) \, ,
\label{eq-phi}
\ee
where $J_B(\phi)$ is the bosonic current (zero-form). In particular,
one has
\be
J_B(\phi) \, = \, 
- \left[ \, H , \left[ H, \phi \right] \, \right] \; , 
\; \; \; \mbox{and} \; \; 
J_B(\phi) \, = \, \frac{1}{2} \,
\left( \phi H^{\dagger} H + H^{\dagger} H \phi \right) 
\label{currents}
\ee
for a Higgs triplet and a Higgs doublet (in matrix
representation), respectively, provided that the latter transforms 
by left multiplication under the action of SU(2).

Equation (\ref{eq-phi}) is the electric part of the YM equation.
The twist Eq. (\ref{eq-a}) assumes the form of a differential
conservation law. This is due to the fact that the connection 
$a$ is an Abelian gauge field which -- for reasons of gauge 
invariance -- enters the effective action only via the field 
strength $da$. All stationary self-gravitating matter models 
give, therefore, rise to a generalized twist potential, 
$\chi$, say. It is well known that the twist potential for 
the Einstein-Maxwell system involves the electric {\it and} 
the magnetic potential. The above reasoning implies that the 
twist potential continues to exist in the EYM system, although 
scalar magnetic potentials cease to do so in non-Abelian gauge 
theories. In fact, Eq. (\ref{eq-a}) implies the existence of a 
function $\chi$, such that
\be
\left( 1 + 4 \kappa \, \sigma^{-1} \Trace{\phi^2} \right) \, da \, = \, 
\sigma^{-2} \ast d \chi \, - \,
4 \kappa \, \sigma^{-1} \Trace{\phi F} \, . 
\label{twist-pot}
\ee
(It may be worthwhile mentioning that an explicit expression
for the twist potential does not exist for a rotating boson star. 
This is a consequence of the fact that the effective action
does contain terms in $a$ itself, since the model is not 
stationary in the strict sense and is, therefore, only
gauge invariant under a combined transformation involving $a$ 
{\it and} the time coordinate.)

\section{Stationary perturbations of static spacetimes}

Let us now consider stationary perturbations of a 
static (i.e., $a = 0$) EYM configuration. The above reasoning
implies that the perturbations $\delta a$ and $\delta \phi$
do {\it not} couple to the remaining metric 
and matter perturbations, provided that the static 
configuration is {\it purely magnetic}.
(In this case both $a$ and $\phi$ are first order quantities.)
The stationary perturbations of a static, purely
magnetic spacetime therefore fall into two complementary
sets, henceforth called {\it static} perturbations and 
{\it purely stationary} perturbations. The static set 
involves only perturbations of fields (metric and matter) 
which are already present in the equilibrium configuration. 
It is obvious that the restriction to perturbations of 
this kind gives rise to a consistent set of first order 
equations. The purely stationary perturbations involve those
fields which vanish for static, purely magnetic configurations. 
It is an interesting consequence of the above KK 
reduction that the purely stationary perturbations form a 
consistent subset as well, that is, the twist 
channel and the electric channel do {\it not} cause 
perturbations of the remaining fields.

It is very intuitive (and will be shown below) that it is
precisely the set of purely stationary perturbations which gives
rise to angular momentum. Hence, we shall now focus on
these perturbations, that is, we consider
\be
\delta \gtens \, = \, 0 \, , \qquad 
\delta \sigma \, = \, 0 \, , \qquad
\delta A \, = \, 0 \, , \qquad
\delta H \, = \, 0 
\label{background}
\ee
and
\be
a \, = \, \delta a \, \qquad
\phi \, = \, \delta \phi \, .
\label{firstorder}
\ee
The arguments presented above imply that the static equations 
for $\gtens$, $\sigma$, $A$ and $H$ remain unchanged in first 
order perturbation theory. The perturbation equations for 
$\delta a$ and $\delta \phi$ are obtained from Eqs. 
(\ref{eq-a}) and (\ref{eq-phi}), respectively. However, it turns 
out to be more convenient to use the linearized 
{\it twist potential}, $\delta \chi$, rather than $\delta a$
itself. The perturbation equation for $\delta \chi$ is derived
from Eq. (\ref{twist-pot}) by linearizing the integrability 
condition $d (da) = 0$, whereas the perturbation equation for
$\delta \chi$ is obtained from Eqs. (\ref{eq-phi}) and 
(\ref{twist-pot}). One easily finds (to first order in
$\delta \chi$ and $\delta \phi$)
\be
- \, \frac{1}{4 \kappa} \, 
d \left( \frac{1}{\sigma^{2}} \ast d \, \delta \chi \right) \, + \,
d \left( \frac{1}{\sigma} \, \Trace{F \, \delta \phi} \right) 
\, = \, 0 \, ,
\label{eq-1}
\ee
\be
D \left( \frac{1}{\sigma} \ast D \delta \phi \right) \, + \,
\frac{1}{\sigma} F \we d \, \delta \chi \, = \, 
4 \kappa \, \Trace{F \, \delta \phi} \we \ast F \, + \, 
\frac{1}{\sigma^{2}} \ast J_B(\delta \phi) \, .
\label{eq-2}
\ee
The above equations for the scalar perturbations 
$\delta \chi$ and $\delta \phi$ form a formally 
{\it self adjoint} system. This is manifest for the second 
order differential operators and for the diagonal potential 
terms on the right hand side of Eq. (\ref{eq-2}). The two 
off-diagonal parts on the left hand sides are easily seen to 
be symmetric as well. Moreover, $\delta \chi$ and $\delta \phi$
are {\it gauge invariant} perturbation amplitudes: This is 
obvious for $\delta \chi$, since it is obtained from the 
Abelian field strength $\delta (da)$. The invariance of 
$\delta \phi$ follows from the infinitesimal transformation 
law $\delta \phi \rightarrow \delta \phi + 
U_{\star}(\phi) \delta f$ and the fact that $\phi$ vanishes 
for the background solution. [We recall that under an 
infinitesimal gauge transformation, $\delta f$, one has 
$\delta \fourA \rightarrow \delta \fourA + \fourD (\delta f)$.]
Before we proceed, we shall briefly argue that the angular 
momentum of a stationary spacetime involves only the purely 
stationary set of perturbations, governed by 
eqs. (\ref{eq-1}) and (\ref{eq-2}).

Apart from stationarity, no symmetry requirements have been
imposed so far. We shall now assume that spacetime admits a second,
axial Killing field, $\partial_{\varphi}$, and compute the 
Komar expression for the angular momentum ${\cal J}$. Asymptotic 
flatness implies that only the terms which are linear in $a$ 
contribute to the Komar integral,
\be
{\cal J} \, = \, \frac{1}{16 \pi \/ G} \,
\int_{S^{2}_{\infty}} \fourast \, d\psi^{(4)} \, = \,
\frac{1}{16 \pi \/ G} \,
\int_{S^{2}_{\infty}} \left[ \, a \we \ast d\psi - 
\psi \we \ast da \, \right] \, .
\label{Komar}
\ee
Here, $\psi^{(4)} = g^{(4)}_{\varphi \mu} dx^{\mu}$ 
is the axial Killing one-form
and $\psi$ its projection on $\Sigma$. Since
in the asymptotic regime $\sigma \rightarrow 1$ and 
$\psi \rightarrow r^2 \sin^2 \! \vartheta d \varphi$,
the first integrand in Eq. (\ref{Komar}) becomes equal to
minus two times the second one. Hence, the angular momentum becomes
\be
{\cal J} \, = \, -\frac{3}{16 \pi \/ G} \, \int_{S^{2}_{\infty}}
\psi \we \ast da \, = \, 
\frac{3}{16 \pi \/ G} \, \int_{S^{2}_{\infty}}
r \/ a \we d (\cos \! \vartheta) \, .
\label{Komar-2}
\ee

Let us now consider arbitrary stationary, axisymmetric 
perturbations of a static and axisymmetric spacetime. 
In this case, $a$ {\it is} a first order quantity, and 
the expression for ${\cal J}$ involves neither 
perturbations of the $3$-metric $\gtens$ nor of the KK 
scalar field $\sigma$. Hence, only the purely stationary 
modes, governed by Eqs. (\ref{eq-1}) and (\ref{eq-2}), 
contribute to the angular momentum. 

\section{Multipole expansion}

We now restrict ourselves to 
spherically symmetric background configurations and perform 
a multipole expansion of the relevant first order quantities 
(which, for simplicity, are assumed to be axisymmetric). 
In the unperturbed spacetime 
($\RR \times \Sigma\,,\,\gtens^{(4)}$), we use standard 
Schwarz\-schild coordinates and parameterize the metric 
$\fourg=-\sigma\,dt^2+\,\sigma^{-1}\,\gtens$ in the familiar 
form 
\be
\sigma=NS^{2}\;,\qquad
\sigma^{-1}\,\gtens \,\, = \,\, N^{-1} dr^{2} \, + \, 
r^{2} \, d \Omega^{2} \, ,
\label{h-metric}
\ee
where $N$ and $S$ are functions of the coordinate $r$.
In the ``canonical gauge'', the static, spherically 
symmetric, purely magnetic background YM potential assumes 
the form 
\be
A \, = \, [\,1-w(r)] \, \hat{\ast} d \tau_r \, ,
\label{back-A}
\ee
where $\hat{\ast}$ denotes the Hodge dual with respect 
to the standard metric on $S^2$, and
$\tau_r,\tau_{\vartheta},\tau_{\varphi}$ are the spherical 
generators of SU(2) (normalized such that 
$[\tau_{\vartheta},\tau_{\varphi}] = \tau_{r}$).
(See also \cite{OB96} for a discussion of symmetric
gauge fields with a higher rank gauge group.)
For a static, spherically symmetric Higgs field we have
\be
H^{(3)} \, = \, h(r) \, \tau_r \, ,
\qquad
H^{(2)} \, = \frac{1}{2} \, h(r) \, \idid \, ,
\label{back-H}
\ee
where, as before, $H^{(3)}$ and $H^{(2)}$ denote a Higgs
field in the adjoint (triplet) and the fundamental (doublet)
representation of SU(2), respectively. (We recall that the 
general spherically symmetric ansatz for a Higgs doublet is 
$H^{(2)} = \frac{1}{2} h(r) \idid -  g(r) \tau_r$, and that
the magnetic gauge potential $A$ involves the additional term 
$\tilde{w}(r) d\tau_r$. However, in the {\em static} case, the 
field equations imply that one may consistently set 
$g(r)=\tilde{w}(r)=0$; see, e.g., \cite{BBMSV}.)

Let us now consider the multipole expansion for the perturbations.
We first observe that the perturbations of the metric potential
$a$ which contribute to the ADM angular momentum belong to the 
sector with (total) angular momentum $j=1$. In fact, as $\delta a$ 
is an axisymmetric one-form on the spherically symmetric 
manifold $\Sigma$, this has an expansion of the form
\be
\delta a \, = \, 
\sum_{j} \; \left[ \,
\alpha_{j} \, \hat{\ast}\,d Y_{j} 
\; + \;
\beta_{j} \, Y_{j}\, dr 
 \; + \;
\gamma_{j} \, d Y_{j} \, \right] \; ,
\label{expansion-deltaa}
\ee
where the coefficients are functions of the 
radial coordinate $r$, and $Y_{j}$ is a short hand 
for the spherical harmonics $Y_{j0}$. Since the integrand 
in the Komar expression (\ref{Komar-2}) is proportional 
to $dY_{1}$, the orthogonality of the spherical 
harmonics implies that only the term  proportional to
$\hat{\ast}\, dY_1$ in the expansion for $\delta a$ 
gives a non-trivial contribution. Hence, as claimed, 
the sector describing infinitesimal rotations consists of the 
purely stationary perturbations 
with total angular momentum $j=1$. 

Next, we evaluate the perturbation equations (\ref{eq-1}),
(\ref{eq-2}) for the background fields $A$ and $H$
(given in eqs. (\ref{back-A}) and (\ref{back-H}), respectively),
which are easily seen to be symmetric under parity. 
To this end, we first expand the electric 
YM perturbation $\delta \phi$ 
in terms of the ``isospin'' harmonics ${C}^{\,\ell}_{j\/m}$, 
which, after suitable identifications, are proportional 
to the standard vector harmonics $Y^{\,\ell}_{j\/m}$:  
\be
C^{\,j}_{j\/m} \, = \, \tau_A \, \varepsilon^{AB}\,\hat{\nabla}_{\!B} 
Y_{jm} \, ,
\qquad 
C^{\,j\pm1}_{j\/m} \, = \, 
\mp\,\frac{1}{2}(2j + 1 \pm 1) \, \tau_r \, Y_{jm}
\; + \;
\tau_A\, \delta^{AB}\,\hat{\nabla}_{\!B}Y_{jm} 
 \;,
\label{isoharmon}
\ee
where capital Latin letters refer to indices with 
respect to the orthonormal frame 
${\theta}^{\vartheta} = d \vartheta$, 
${\theta}^{\varphi} = \sin \! \vartheta\, d \varphi$ 
on $S^2$. The harmonics $C^{\,\ell}_{j\/m}$ have 
parity $(-1)^{\ell}$ and are, of course, 
eigenfunctions of the Laplacian 
$\hat{\Delta} = \hat{\ast} d \hat{\ast} d$ on $S^2$
with eigenvalues $-\ell(\ell+1)$. 
It is not hard to see that the symmetry under a parity 
transformation implies that the odd parity component of 
$\delta \phi$ decouples. Moreover, this does not contribute to 
the ADM angular momentum, since the parity of the corresponding 
variation of $a$ is also odd [see Eq. (\ref{twist-pot})]. 
Thus, the axial perturbations which are relevant to 
infinitesimal rotations can be parameterized in terms of 
three scalar functions $\bx(r)$, $\by(r)$ and $\bz(r)$:
\be
\delta \chi \, = \, \sqrt{2\kappa} \, \bx(r) \, Y_1 \, ,
\qquad 
\delta \phi \, = \, \by(r) \, \tau_r Y_1 \, + \, 
\bz(r) \, \frac{1}{\sqrt{2}} \,
\tau_{\vartheta}\, \partial_{\vartheta}\,Y_1 \, .
\label{expansion}
\ee
 
At this point, it is a straightforward task to derive 
the perturbation equations for the vector-valued function 
$\underline{\bv}=(\bx,\by,\bz)^T$ from Eqs. (\ref{eq-1}) 
and (\ref{eq-2}). The rotational deviations are 
governed by the following Sturm-Liouville equation:
\be
\left\{-\,\partial\, r^2 \bbAA \,
\partial \, + \,\bbJJ \, + \, \bbBB \,\partial\, -
\, \partial\, \bbBB^T \,+\, \bbPP \right\}
\underline{\bv} \, = \, 0 \,\,,
\label{perturb-eqs}
\ee
where $\partial$ denotes the differential operator
\be
\partial f \, \equiv \, f' \, \equiv \, 
\frac{1}{S} \, \frac{d f}{d r} \, ,
\label{der}
\ee
and $S$ is defined in Eq. (\ref{h-metric}).
The first two terms originate from
the differential operators
$D \left( \sigma^{-1}\! \ast\! D \delta \phi \right)$ and
$d \left( \sigma^{-2}\! \ast\! d \delta \chi \right)$, which 
give rise to the matrix-valued background functions
\be
\bbAA \, =  
\left( \begin{array}{ccc}
-\sigma^{-1} & 0 & 0 \\
0 & 1 & 0 \\
0 & 0 & 1 
\end{array} \right) \, ,\qquad
\bbJJ =\, \frac{1}{\sigma}
\left( \begin{array}{ccc}
-2\sigma^{-1}&0 & 0\\
0& 2 (w^2 + 1) & -2 \sqrt{2} w\\
0&-2 \sqrt{2} w & w^2 + 1
\end{array} \right) \, .
\label{D-and-J}
\ee
(Note that for $w\to 1$ and $\sigma\to 1$ the 
eigenvalues of $\bbJJ$ become $-2$, $0$, $6$, which 
reflects the fact that the twist channel has angular 
momentum $j=\ell=1$, whereas the orbital angular momentum 
of the YM perturbations is $0$ and $2$.) 
For the differential coupling between the twist potential 
and the gauge fields we obtain 
(in units with $\kappa/2=4\pi G/g^2=1$)
\be
\bbBB\,\partial\, -\, \partial\, \bbBB^T
\ = \,2\,\left( \begin{array}{ccc}
0 &  \, \partial \, \sigma^{-1}({w^2 - 1})\, &0 \\
-\sigma^{-1}({w^2 - 1}) \, \partial & 
0 & 0\\
0 & 0 & 0
\end{array} \right) \, .
\label{PP0}
\ee
Finally, the potential matrix $\bbPP$ is given by
\be
\bbPP \, =\, -\,\frac{2}{\sigma}\left( \begin{array}{ccc}
0 & 0 &  \sqrt{2} \, {w'}\\
0 &  2 \,{(w^2-1)^2}{r^{-2}} &
0 \\
 \sqrt{2} \, {w'} & 
0 & 2 \, {\sigma} \, w'^2
\end{array} \right) 
+ \; \bbPP_h \; ,
\label{PP}
\ee
where the background Higgs field enters the perturbation
equations only via the matrix $\bbPP_h$, which becomes
\be
\bbPP_h^{(3)} \, = \, \frac{r^2}{\sigma}
\left( \begin{array}{ccc}
0 & 0 &  0\\
0 & 0 &  0 \\
0 & 0 &  h^2
\end{array} \right)
\; , \; \; \mbox{and} \; \; \;
\bbPP_h^{(2)} \, = \, \frac{r^2}{4 \/ \sigma}
\left( \begin{array}{ccc}
0 & 0 &  0\\
0 & h^2 &  0 \\
0 & 0 &  h^2
\end{array} \right)
\; ,
\label{Higgsmatrices}
\ee
for a Higgs triplet and a Higgs doublet, respectively. 

In order to discuss the pulsation equations one needs
the behavior of the background 
quantities $N$, $S$, $w$ and $h$. These are subject
to the static, spherically symmetric EYM-Higgs equations, 
which are most conveniently obtained from the effective
Lagrangian. For the gravitational part one finds 
(up to an exact differential)
$\fourast \fourR = 4 S \frac{dm}{dr} dt \we dr \we d\Omega$,
where $2m(r) = r[1-N(r)]$; see, e.g. \cite{BHS96}.
Also evaluating the effective Lagrangians 
(\ref{L-YM}) and (\ref{L-H}) [with $a = 0$ and $\phi = 0$] 
immediately gives the static, spherically symmetric action
(using again $\kappa/2 = 1$)
\be
{\cal S} \, = \, 
\frac{1}{G} \left(
-\frac{dm}{dr} + N \left(\frac{dw}{dr}\right)^2 + 
N \frac{r^2}{2} \left(\frac{dh}{dr}\right)^2 + 
\frac{(w^2 - 1)^2}{2 \, r^2} + \frac{r^2}{2} P(h) + Q(w,h) \right)
\, S \, dr \, ,
\label{sph-symm-lag}
\ee
where $P(h)$ denotes the Higgs potential, and 
the interaction potential $Q(w,h)$ is given by
\be
Q^{(3)}(w,h) \, = \, h^2 \, w^2 
\; , \; \; \mbox{and} \; \; \;
Q^{(2)}(w,h) \, = \, \frac{1}{4} \, h^2 \, (1-w)^2 \, , 
\label{w-h-pot}
\ee
for a Higgs triplet and a Higgs doublet, respectively.
Variation of ${\cal S}$ with respect to $m$ and $S$ yields
the relevant Einstein equations, whereas variation
with respect to $w$ and $h$ gives the magnetic YM-Higgs 
equations. Using the background equations enables one now 
to analyze the perturbation equations in the vicinity 
of the origin, the horizon and in the asymptotic regime.
In the following section we present the results of a systematic
discussion.

\section{Rotating black holes}

We start by discussing the behavior of perturbations near the 
horizon, $r_H$, of a given black hole background. If the 
unperturbed solutions are analytic in a neighborhood of the 
horizon, then $r_H$ is a regular singular point of the 
perturbation equations. Local properties of the solutions 
can, therefore, be analyzed by means of standard techniques. 
In particular, the number of physically acceptable solutions 
is easily determined: The perturbation equations for the EYM 
system coupled to a Higgs doublet or a Higgs triplet admit 
precisely {\it four\/} independent solutions which are admissible 
near the horizon (provided that the unperturbed black hole is 
not extreme).

Next we consider the asymptotic regime, $r\to\infty$. Near 
infinity, the background solutions with a Higgs field in the
adjoint representation approach the embedded
Reissner-Nordstr\"om solution with magnetic charge $P^2=1$: 
$w\approx 0$, and $|h|\approx v$, where $v$ is the vacuum 
expectation value of the Higgs field.  Similarly, the
unperturbed solutions with a Higgs field in the fundamental 
representation approach the embedded Schwarzschild solution: 
$|w| \approx 1$, and $|h|\approx v$. (The Abelian nature of 
the matter fields becomes manifest after a suitable 
gauge transformation.) It is straightforward to verify that 
the leading asymptotic behavior of the perturbations remains 
unchanged if a given background solution is replaced by 
its ``asymptotic Abelian part''. Within this approximation, 
the perturbation equations simplify considerably in the asymptotic
regime: For a Higgs triplet, the ``massive'' perturbation channel 
decouples, and the remaining two equations have a regular 
singular point at infinity. For a Higgs doublet, the asymptotic 
system can even be decoupled completely. For both types of 
Higgs fields it is, therefore, 
readily verified that precisely {\it three\/} independent 
solutions exist which are physically acceptable near infinity.

Since the background configurations are continuous 
for $r_H<r<\infty$, the above-defined local solutions have 
extensions with a range of definition containing the whole 
interval $r_H<r<\infty$. By construction, these extensions 
span the subspaces of global solutions which are acceptable 
near the inner and the outer boundary point, respectively. 
Since these solution-subspaces have dimension 
{\/three\/} and {\/four\/}, respectively, and since the 
dimension of the total solution-space is {\/six\/}, the 
intersection of the subspaces is 
(at least) {\/one\/}-dimensional. Thus, physically acceptable
global solutions of the perturbation equations 
{\/always\/} exist for the EYM-Higgs system.

\section{Rotating solitons}

Like in the black hole case, the perturbation equations 
for soliton background solutions have a regular singular 
point at the inner boundary point, $r=0$, provided that 
the unperturbed solutions are analytic in a 
neighborhood of the origin. In the vicinity of this point, 
the leading behavior of perturbations is completely fixed 
by the ``centrifugal barrier'', ${\cal J}/r^2$. It is, 
therefore, straightforward to verify that precisely 
{\it three\/} independent solutions exist which are globally 
defined and physically acceptable near the origin. In the 
asymptotic regime, $r\to\infty$, the behavior of 
perturbations is the same as in the black hole case. 
Hence, the global solutions of the perturbation equation 
which are admissible near both boundary points are given 
by the intersection of two solution-subspaces, each of
which is {\it three\/}-dimensional. Since the intersection 
of two three-dimensional subspaces of a six-dimensional 
linear space generically is trivial, we are led to the 
conclusion that soliton solutions of the EYM-Higgs system 
generically do not admit rotational excitations.  

\section{Concluding remarks}

Both, the general structure and the main features of 
the perturbation equations are dominated by the EYM part 
of the system. It is, therefore, natural to expect 
that the above results, derived for the SU(2) EYM-Higgs system, 
continue to hold for a class of EYM systems with higher 
rank gauge groups, and more general bosonic matter fields.
Hence, we conjecture that bosonic EYM black holes always have 
rotating counterparts, whereas bosonic EYM solitons
generically do not admit infinitesimal rotations. The approach 
presented in this paper offers the possibility for a 
{\it systematic\/} study of these conjectures, which, in case 
they should turn out to be correct, raise the important question 
about the physical mechanism preventing bosonic
solitons from rotating.

\section{Acknowledgments}

It is a pleasure to thank Norbert Straumann and Michael Volkov 
for stimulating discussions. This work was supported by the Swiss 
National Science Foundation.

\end{document}